\documentstyle[eqsecnum,preprint,aps,prd]{revtex}
\begin{document}
\draft
\preprint{SUSX-TH/96-013 hep-ph/9609510}
\title{The Dynamical Nonabelian Two-Form: BRST Quantization}
\author{\bf Amitabha Lahiri}
\address{Centre for Theoretical Physics, \\ 
University of Sussex, Brighton BN1 9QH. U. K.\thanks{Till Sept. 27 
1996}}
\address{mpfq3@pcss.maps.susx.ac.uk}
\date{September, 1996}
\maketitle
\begin{abstract}
When an antisymmetric tensor potential is coupled to the field strength 
of a gauge field via a $B\wedge F$ coupling and a kinetic term for $B$ 
is included, the gauge field develops an effective mass. The theory can 
be made invariant under a non-abelian vector gauge symmetry by  
introducing an auxiliary vector field. The covariant quantization of 
this theory requires ghosts for ghosts. The resultant theory including 
gauge fixing and ghost terms is BRST-invariant by construction, and 
therefore unitary. The construction of the BRST-invariant action is 
given for both abelian and non-abelian models of mass generation.

\end{abstract}
\medskip

\section{Introduction}

The free antisymmetric tensor potential has one degree of freedom, a
scalar\cite{opd}.  This scalar can be coupled to an abelian gauge field
via a `topological' $B\wedge F$ term with a dimensionful coupling
constant $m$ of mass dimension one.  The resulting theory, which is
classically dual to the Goldstone model (the abelian St\" uckelberg
model), has three degrees of freedom which can be identified, both
classically and quantum mechanically, with the propagating degrees of a
massive gauge field of mass $m$\cite{aurtak,trg,abl,minwar}.  This
theory, as well as its vacuum, is invariant under both $U(1)$ and the
vector gauge symmetry $B_{\mu\nu} \to B_{\mu\nu} +
\partial_{[\mu}\Lambda_{\nu]}$ with an arbitrary vector field
$\Lambda_\mu$.  In other words, this model generates vector boson
masses without symmetry breaking and without a residual Higgs.  The
symmetries of the theory ensure that when fermions are included in the
theory, only the transverse components of the gauge field couples to
the fermionic current.  The generic coupling term of mass dimension
four between the antisymmetric tensor and fermions is of the form
$\bar\psi(a + b\gamma_5)\sigma^{\mu\nu}B_{\mu\nu}\psi$, which is not
invariant under the vactor gauge transformations, and therefore cannot
be included in the action if this symmetry is to be maintained.  This
implies that there is no three-point coupling, and therefore no loop,
directly involving $B_{\mu\nu}$.  Consequently it is straightforward to
renormalize QED in which photons acquire mass via 
this mechanism \cite{abl}.

The possibility that a non-abelian version of this theory may exist as
a consistently quantizable theory is an interesting one.  Although many
aspects of the Standard Model have been experimentally verified, the
symmetry-breaking sector is still mostly unexplored and the source of
some unanswered questions.  So far experiments have not turned up an
elementary scalar in any system of interacting particles, nor is there
any positive evidence of an electroweak Higgs particle, either
elementary or composite, at currently available energies.  On the other
hand, various theoretical arguments set the upper bound of the Higgs
mass only a little out of reach of the present generation of
accelerators.  This suggests that perhaps we should consider
alternative descriptions of the symmetry-breaking sector of the
electroweak theory and prepare ourselves for the situation that no
Higgs particle is ever found.

The Higgs sector as it stands has three equally important roles.  One is
to break the global $SU(2)_{isospin}\times U(1)_{hypercharge}$ symmetry
down to the $U(1)$ symmetry of electromagnetism.  In the Standard Model
the mechanism of symmetry breaking generates masses for the vector
bosons $W^{\pm}$ and $Z$.  In addition, the Yukawa coupling of the Higgs
scalar to fermions breaks chiral symmetry and contributes to fermion
mass generation.  But suppose we consider the possibility that the three
questions may be resolved separately.  Then it makes sense to consider a
mechanism to generate masses for vector bosons via a $B\wedge F$
interaction with an antisymmetric tensor, and look for the possibility
of symmetry breaking and fermion mass generation in some other
interaction in the theory, possibly as dynamical mechanisms.

But first we have to have a theory that can be consistently quantized,
i.e., one that is both unitary and renormalizable.  Various Higgs-free
theories of massive non-abelian vector bosons, including the Proca
model, the St\"uckelberg model, the gauged non-linear sigma model, or
the Higgs model with a heavy Higgs, are either nonrenormalizable or
violate unitarity.  Therefore any other proposed mechanism must pass
these two tests.  As far as the antisymmetric tensor is concerned the
renormalizability of the abelian theory does not really provide a
pointer, because even a gauge variant mass term for the photon does not
affect the renormalizability of QED \cite{collins}.  However, as was
pointed out elsewhere \cite{gvm}, it is possible to construct a
non-abelian theory which is power-counting renormalizable, has unbroken
gauge symmetries, and has propagators which fall off as $1/k^2$ at high
momentum, so there are no obvious obstructions to renormalizability.
(Unlike the Freedman-Townsend model \cite{fretow} which does not have a
kinetic term for $B_{\mu\nu}$, the model proposed in \cite{gvm} is not
dual to the non-linear sigma model.)  But unitarity is another story.

The biggest argument faced by any theory with massive vector bosons but
without a Higgs-like excitation involves unitarity.  Any theory with a
hermitian Hamiltonian operator is necessarily unitary.  However, a
gauge theory has several redundant degrees of freedom which have to be
eliminated by gauge fixing.  An explicitly Lorentz-covariant gauge
fixing term introduces states of negative norm in the theory which have
to be eliminated in turn by introducing ghost fields.  At this point
the theory contains non-hermitian fields and states of negative norm,
so the unitarity of the theory needs to be checked explicitly.  One way
of checking whether a theory unitary is to see if the action including
the gauge fixing and ghost terms is invariant under BRST
transformations \cite{brs,tyu}.  If it is invariant, it is possible to
define the conserved Noether charge $Q$ of the symmetry.  This charge
is nilpotent, $Q^2 = 0$, and defines a cohomology on the Fock space of
the theory.  The space of states $|\psi\rangle$ such that
$Q|\psi\rangle = 0 $ but $|\psi\rangle \neq Q|\chi\rangle$ for any
$|\chi\rangle$ can be identified with the physical subspace of the Fock
space, and it can be shown that the S-matrix of the theory in is
unitary in this physical subspace \cite{bec}.

For the antisymmetric tensor potential, the Fadeev-Popov construction
runs into problems because of the need for ghosts for ghosts
\cite{thibau}.  It is well known that the constraints of the free
antisymmetric tensor form a reducible system \cite{hentei}, as do the
constraints of the pure $B\wedge F$ action.  What is not so obvious (or
well known) is that the constraints form a reducible system, both in
the abelian and the non-abelian models, even when both the kinetic term
and the $B\wedge F$ coupling term are present in the action
\cite{abcon,nabcon}.  (This is just a restatement of the fact that it
is possible to introduce a kinetic term for $B_{\mu\nu}$ without
breaking the vector gauge symmetry, and without introducing extra
degrees of freedom.)  As a result, ghost-for-ghosts are still a
necessity, which causes problems for the Fadeev-Popov construction.  A
long time ago a geometric construction was proposed \cite{thibau} for
the construction of the BRST-anti-BRST-invariant quantum action for the
Freedman-Townsend model.  More recently, a geometric construction was
proposed using a similar `horizontality condition' \cite{hwalee} for
the model of vector boson mass generation with a non-abelian
antisymmetric tensor.  A BRST-anti-BRST-invariant action was found this
way. Therefore it is known that a covariant gauge fixed quantum action 
exists for the mass generation mechanism.

In this paper I demonstrate that it is possible to construct a 
BRST-invariant tree-level action in a covariant gauge starting from the 
classical action proposed in \cite{gvm} and proceeding in a similar 
fashion to the textbook construction \cite{collins} for the free 
Yang-Mills theory. In section 2,  the BRST-invariant action for the 
abelian model is constructed, both for the sake of completeness and as 
a test case. The BRST transformations of the various fields and their 
ghosts in the non-abelian model can be intuited from the abelian case. 
In section 3, the BRST transformations of the non-abelian fields are 
given, following as closely as possible the constructions for the 
abelian model and the free Yang-Mills theory. Section 4 contains a 
summary and discussion of results.

\section{The Abelian Model}

Let me begin by discussing the construction of a BRST-invariant quantum 
effective action for the dynamical abelian two-form coupled to a gauge 
field. The theory under consideration is described by the classical 
action
\begin{equation}
S_0 =
\int d^4x\big(-{1\over4}F_{\mu\nu}F^{\mu\nu} - 
{1\over12}H_{\mu\nu\lambda} H^{\mu\nu\lambda} 
 + 
{m\over4}\epsilon^{{\mu\nu\lambda}\rho}F_{\mu\nu}B_{\rho\lambda}\big).
\label{abac}
\end{equation}
where $F_{\mu\nu}$ and $H_{\mu\nu\lambda}$ are the respective field 
strengths of 
$A$ 
and $B$, $F_{\mu\nu} = \partial_{[\mu}A_{\nu]} = \partial_\mu A_\nu - 
\partial_\nu 
A_\mu$ and
$H_{{\mu\nu\lambda}} = \partial_{[\mu}B_{\nu\lambda]}
\partial_\mu B_{\nu\lambda} + \partial_\nu B_{\lambda\mu} + 
\partial_\lambda
B_{\mu\nu}$. This action 
remains
invariant under the independent gauge transformations
\begin{eqnarray}
A_\mu &\to& A_\mu + \partial_\mu\chi, \qquad B_{\mu\nu} \to
B_{\mu\nu},\label{abgsym} \\
A_\mu &\to& A_\mu,\qquad
B_{\mu\nu} \to B_{\mu\nu} + \partial_{[\mu}\Lambda_{\nu]}.
\label{abksym}
\end{eqnarray}

This theory has three degrees of freedom \cite{abcon}, one of which
couples to $A_\mu$ in a fashion similar to the Goldstone mode in the
Higgs mechanism.  The interaction between the gauge field and the
antisymmetric tensor has a two-point vertex operator proportional to
the momentum.  The `physical' propagator --- so called because it
couples to external fermion currents --- can be calculated by summing
over all gauge propapgators with insertions of antisymmetric tensor
propagators \cite{abl}.  The physical propagator has a pole at $k^2 =
m^2$, i.e.  this theory can be thought of as a (gauge-invariant) theory
of a massive abelian gauge field, with no other degree of freedom.

In this section I shall give a straightforward construction of
the BRST-invariant action for the Abelian model (\ref{abac}).  Starting
with the free action $S_0$, the gauge-fixing terms in the covariant
Lorentz gauge are added, and the Fadeev-Popov ghost terms are computed
so as to exactly cancel the variation of the gauge fixing terms.  The
notation used in this section and the next one follows that of
\cite{collins}.  The BRST transformations of $A_\mu$ and $B_{\mu\nu}$
are given by their gauge transformations with grassmann-valued gauge
parameters $\omega$ and $\omega_\mu$ respectively,
\begin{equation} 
\delta A_\mu =
\partial_\mu\omega {\delta\lambda} ;\qquad \delta B_{\mu\nu} = 
(\partial_\mu\omega_\nu -
\partial_\nu\omega_\mu) {\delta\lambda}.  
\label{abfbrs} 
\end{equation} 
As is obvious, there is
a further symmetry under which $\omega_\mu$ is shifted by the gradient
of a scalar.  This implies that the effective action needs to be
gauge-fixed for $\omega_\mu$ as well, otherwise the ghost propagator
does not exist.  This introduces a commuting ghost $\beta$ for
$\omega_\mu$.  I can now choose the gauge-fixing part of the effective 
action to be 
\begin{equation} 
{\cal L}_{gf} = - {1\over{2\xi}}(F_1)^2 -
{1\over{2\eta}}F_2^\mu F_{2\mu} - {1\over{2\zeta}}(F_3);
\end{equation}
where the fields are fixed in covariant gauges,
\begin{equation}
F_1 = \partial_\mu A^\mu,\quad F_2^\mu = \partial_\nu B^{\mu\nu}, 
\quad F_3 =
(\partial_\mu\bar{\omega}^\mu)(\partial_\nu\omega^\nu).  
\label{ablgf} 
\end{equation}

The BRST transformations of the ghost fields can now be written down 
along the lines of the standard procedure for gauge theories, 
\begin{eqnarray}
\delta\omega & = & 0,\nonumber\\
\delta\bar{\omega} & = & {1\over\xi}\partial_\mu 
A^\mu{\delta\lambda},\nonumber\\
\delta\omega_\mu & = & \partial_\mu\beta{\delta\lambda},\nonumber\\
\delta\bar{\omega}_\mu & = & {1\over\eta}\partial^\nu 
B_{\mu\nu}{\delta\lambda},\nonumber\\
\delta\beta & = & 0,\nonumber\\
\delta\bar{\beta} & = & - 
{1\over\zeta}(\partial_\mu\bar{\omega}^\mu){\delta\lambda}.
\label{abgbrs}
\end{eqnarray}
The ghost terms in the action are chosen to compensate for the 
variation in the gauge-fixing terms, and are therefore
\begin{equation}
{\cal L}_{FP} = \partial_\mu\bar{\omega}\partial^\mu\omega - 
\partial_\mu\bar{\omega}_\nu(\partial^\mu\omega^\nu - 
\partial^\nu\omega^\mu) + 
\partial_\mu\bar{\beta}\partial^\mu\beta.
\label{ablgc}
\end{equation}
The total action,
\begin{equation}
S = S_0 + \int d^4x{\cal L}_{gf} + \int d^4x{\cal L}_{FP},
\end{equation}
is now fully gauge-fixed but is invariant under the BRST 
transformations as given in (\ref{abgbrs}).

Under a BRST transformation the variation in the action can be written 
as a total divergence, 
\begin{eqnarray}
\delta S & = & 
\int \partial_\mu Y^\mu = 0,\nonumber\\
Y^\mu & = & {m\over 2}\epsilon^{{\mu\nu\lambda}\rho}\omega_\nu 
F_{\lambda\rho} - 
{1\over\xi}(\partial_\nu A^\nu)\partial^\mu\omega \nonumber\\
&& + {1\over\eta}(\partial^\lambda
B_{\nu\lambda})(\partial^\mu\omega^\nu - \partial^\nu\omega^\mu) - 
{1\over 
\zeta}(\partial_\nu\bar{\omega}^\nu)\partial^\mu\beta.
\end{eqnarray}
The conserved Noether current for the BRST symmetry is thus
\begin{eqnarray}
j^\mu &=& \sum{{\delta {\cal 
L}}\over{\delta\partial_\mu\phi}}{{\delta\phi}
\over{{\delta\lambda}}} - Y^\mu \nonumber\\ &=& - 
F^{\mu\nu}\partial_\nu\omega + {m 
\over 2}\epsilon^{{\mu\nu\lambda}\rho}\partial_\nu\omega 
B_{\lambda\rho} - 
{1\over\xi}(\partial_\nu A^\nu)\partial^\mu\omega \nonumber\\ && - 
(\partial^\mu\bar{\omega}^\nu -
\partial^\nu\bar{\omega}^\mu)\partial_\nu\beta - {1\over 
2}H^{\mu\nu\lambda}(\partial_\nu\omega_\lambda - 
\partial_\lambda\omega_\nu) \nonumber\\ 
&& + {1\over\eta}(\partial^\sigma B_{\nu\sigma})(\partial^\mu\omega^\nu 
- 
\partial^\nu\omega^\mu) + 
{1\over{\zeta\eta}}(\partial_\nu\omega^\nu)(\partial_\lambda 
B^{\mu\lambda}) 
\nonumber\\ && - 
{1\over\zeta}(\partial_\nu\bar{\omega}^\nu)\partial^\mu\beta - 
{m\over 2}\epsilon^{{\mu\nu\lambda}\rho}\omega_\nu F_{\lambda\rho}.
\end{eqnarray} 
The BRST charge constructed from this current, $Q_{BRST} = \int 
j^0\ d^3x$, is nilpotent, $Q^2_{BRST} = 0$. More explicitly, 
\begin{equation}
{\delta^2\over{\delta\lambda}^2}\left\{ A_\mu, B_{\mu\nu}, \omega, 
\omega_\mu, 
\beta, \bar{\omega}, \bar{\omega}_\mu, \bar{\beta}\right\} = 0,
\label{absqnil}
\end{equation}
where the last three fields satisfy the equality on shell, as is the 
case with $\bar{\omega}$ in free Maxwell theory. Off shell their third 
variations vanish, 
\begin{equation}
{\delta^3\over{\delta\lambda}^3}\left\{ \bar{\omega}, \bar{\omega}_\mu, 
\bar{\beta}\right\} 
= 0.
\label{abcunil}
\end{equation}

\section{The Non-Abelian Model}

The non-abelian model \cite{gvm} starts with a na\"\i ve 
non-abelianization of the action (\ref{abac}) to a compact gauge group, 
which I shall choose to be $SU(N)$ for convenience. To begin with, the
field strength $F_{\mu\nu}$ is now defined as the curvature of an 
$SU(N)$ gauge connection,
\begin{equation}
F_{\mu\nu}^a = (-{i\over g}[D_\mu, D_\nu])^a = \partial_\mu A_\nu^a - 
\partial_\nu A_\mu^a 
- gf^{abc}A_\mu^b A_\nu^c.
\label{Fdef}
\end{equation}
In order to keep the $B\wedge F$ term invariant under $SU(N)$ gauge 
transformations, $B_{\mu\nu}$ has to transform in the adjoint 
representation of the gauge group. This implies that in the kinetic 
term for $B_{\mu\nu}$ the derivative  operator $\partial_\mu$ should be 
replaced by the gauge covariant derivative operator $D_\mu$, and the 
field strength $H_{\mu\nu\lambda}$ should be defined as 
$H_{\mu\nu\lambda} = 
D_{[\mu}B_{\nu\lambda]}.$ The resulting action
\begin{equation}
S = \int d^4x \bigg(- {1\over 4}F_{\mu\nu}^aF^{a\mu\nu} - {1\over 
12}H_{\mu\nu\lambda}^aH^{a\mu\nu\lambda} + {m\over
4}\epsilon^{\mu\nu\rho\lambda}B_{\mu\nu}^aF_{\lambda\rho}^a\bigg), 
\label{nabac1}
\end{equation}
is invariant under $SU(N)$ gauge transformation, but does not contain a 
natural generalization of the vector gauge symmetry (\ref{abksym}) 
under 
which one expects to find $B_{\mu\nu} \to B_{\mu\nu} + 
D_{[\mu}\Lambda_{\nu]}$, with $\Lambda_\mu$ an arbitrary vector field 
transforming homogeneously under the gauge group. Even though this 
is a symmetry of the last term of the action, the second term is not 
invariant under this transformation. The absence of this symmetry shows 
up starkly when one tries to find the propagating degrees of freedom in 
this theory by restricting the fields to the constraint surface 
according to Dirac's prescription. The matrix of Poisson Brackets of 
the constraints turn out to be field-dependent. As a result, it is not 
possible to find local coordinates of the reduced phase space, or a 
Hamiltonian that keeps the degrees of freedom on the constraint 
surface. A detailed analysis of constraints will be presented elsewhere 
\cite{nabcon}, but it turns out that the simplest way to construct a 
reduced phase space is to introduce an auxiliary vector field $C_\mu$, 
also transforming in the adjoint representation of the gauge group, so 
as to compensate for the variation of the action (\ref{nabac1}) under 
the non-abelian vector gauge symmetry. This does not introduce any new 
propagating degrees of freedom, as $C_\mu$ turns out to be fully 
constrained. The need for this auxiliary field also shows up in the 
covariant quantization of the Freedman-Townsend model \cite{thibau}, 
but here its essential purpose \cite{nabcon} is to enforce the 
constraint $[F_{\nu\lambda}, H^{\mu\nu\lambda}] = 0$.

Let me therefore define the compensated field strength $\widetilde 
H_{\mu\nu\lambda}$,
\begin{eqnarray}
\widetilde H_{\mu\nu\lambda}^a &=& (D_{[\mu}B_{\nu\lambda]})^a - 
ig\left[ 
F_{[\mu\nu}, C_{\lambda]}\right]^a \nonumber\\
&=& \partial_{[\mu}B_{\nu\lambda]}^a - 
gf^{abc}A^b_{[\mu}B^c_{\nu\lambda]} 
+ gf^{abc}F_{[\mu\nu}^bC_{\lambda]}^c.
\label{hdef}
\end{eqnarray}
As is obvious, this field strength is invariant under the combined 
transformations
\begin{equation}
B_{\mu\nu} \to B_{\mu\nu} + D_{[\mu}\Lambda_{\nu]}, \quad C_\mu \to 
C_\mu + \Lambda_\mu,
\label{nabkr}
\end{equation}
where $\Lambda_\mu^a$ are real vector fields. It should also be noted 
that the last term in the definition of $\widetilde H_{\mu\nu\lambda}$ 
vanishes in 
the case where the gauge group is abelian, so that $\widetilde 
H_{\mu\nu\lambda}$ 
is an allowed generalization of the abelian field strength. 
 Now I can write down an action which is invariant both under the gauge 
group and the vector transformations (\ref{nabkr}),
\begin{equation}
S_0 = \int d^4x \bigg(- {1\over 4}F_{\mu\nu}^aF^{a\mu\nu} - {1\over 
12}\widetilde 
H_{\mu\nu\lambda}^a\widetilde H^{a\mu\nu\lambda} + 
{m\over
4}\epsilon^{\mu\nu\rho\lambda}B_{\mu\nu}^aF_{\lambda\rho}^a\bigg).
\label{nabac2}
\end{equation}
It should be noted that this action is invariant under the nonabelian
vector gauge symmetry (\ref{nabkr}) without any modification of the
interaction term as long as the fields vanish sufficiently rapidly at
infinity.  Also, the auxiliary field $C_\mu$ is non-dynamical --- there
is no quadratic term corresponding to it in the action, and the
propagator is zero at tree level.  From now on I shall work only with
the compensated field strength $\widetilde H_{\mu\nu\lambda}$ and not
refer to the na\"\i ve field strength $H_{\mu\nu\lambda}$, so I can
drop the tilde and write $H_{\mu\nu\lambda}$ whenever I mean
$\widetilde H_{\mu\nu\lambda}$.

It can be shown by an analysis of constraints that there are three 
degrees of freedom for each gauge index in this theory. The quadratic 
terms in this theory are identical, for each gauge index, to the 
abelian action. As a result, the tree-level effective propagator for 
the gauge field can be computed exactly in the same fashion and leads 
to a pole at $k^2 = m^2$. And there is no residual scalar.

The construction of the BRST-invariant action will follow those for the 
abelian model above and Yang-Mills theory, and also that for the 
pure $B\wedge F$ topological field theory. The gauge-fixing terms are 
easy to write down,
\begin{equation}
{\cal L}_{gf} = - {1\over {2\xi}}(\partial_\mu A^{a\mu})^2 - 
{1\over{2\eta}}(\partial_\nu B^{a\mu\nu})^2 - 
{1\over\zeta}(\partial_\mu\bar{\omega}^{a\mu})(\partial_\nu\omega^{a\nu
}),
\label{nabgf}
\end{equation}
as are the Fadeev-Popov ghost terms,
\begin{equation}
{\cal L}_{FP} = - 
\bar{\omega}^a{\delta\over{\delta\lambda}}(\partial_\mu A^{a\mu}) - 
\bar{\omega}^{a\mu}{\delta\over{\delta\lambda}}(\partial^\nu 
B^a_{\mu\nu}) - 
\bar{\beta}^a{\delta\over{\delta\lambda}}(\partial_\nu\omega^{a\nu}).
\label{nabfp}
\end{equation}
These terms were written down simply by generalizing the abelian case,
and the ghost fields are also defined as generalizations of the abelian
model.  Now, however, an interesting difference shows up.  The fields
$\beta, \bar{\beta}$ were needed in the abelian case in order to
compensate for the gauge fixing of the ghost $\omega_\mu$.  In the
non-abelian model, $\omega^a_\mu$ needs a gauge fixing term for the
same reason, namely that the propagator cannot be defined until that
has been done.  In the abelian model, this showed up as the symmetry of
the action under $\omega_\mu \to \omega_\mu + \partial_\mu\theta$.
Alternatively, the need for this ghost of ghost was a consequence of a
symmetry $\Lambda_\mu \to \Lambda_\mu + \partial_\mu\chi$, with $\chi$
an arbitrary scalar, which is hidden in the vector gauge transformation
(\ref{abksym}).  In the non-abelian model, it is still not possible to
define the ghost propagator and the ghosts need gauge fixing.  One can
try to implement a similar symmetry transformation, $\Lambda_\mu \to
\Lambda_\mu + D_\mu\chi$, where $\Lambda_\mu$ and $\chi$ are now in the
adjoint representation of the gauge group.  However, this leads to the
following set of transformations,
\begin{equation}
\delta B^a_{\mu\nu} = -gf^{abc}F^b_{\mu\nu}\chi^c, \qquad \delta 
C^a_\mu = (D_\mu\chi)^a,
\label{krscalar}
\end{equation}
unlike in the abelian case, where $\delta B_{\mu\nu} = 0$ under such a 
transformation. This implies that there has to be a ghost field 
corresponding to this transformation, as was found by the authors of  
\cite{thibau} in the context of the Freedman-Townsend model. The 
complete set of BRST transformations can now be written down, simply by 
generalizing the abelian case, remembering that all the fields and the 
ghosts transform in the adjoint representation, and including this 
extra ghost,
\begin{eqnarray}
\delta A^a_\mu &=& (D_\mu\omega)^a{\delta\lambda}\nonumber\\
\delta\omega^a &=& - {1\over 2} 
gf^{abc}\omega^b\omega^c{\delta\lambda}\nonumber\\
\delta\bar{\omega}^a &=& {1\over\xi}(\partial_\mu 
A^{a\mu}){\delta\lambda}\nonumber\\
\delta B^a_{\mu\nu} &=& \left( - gf^{abc}B^b_{\mu\nu}\omega^c + 
(D_{[\mu}\omega_{\nu]})^a 
- gf^{abc}F^b_{\mu\nu}\theta^c\right){\delta\lambda}\nonumber\\
\delta C^a_\mu &=& \left( - gf^{abc}C^b_\mu\omega^c + \omega^a_\mu + 
(D_\mu\theta)^a\right){\delta\lambda}\nonumber\\
\delta\omega^a_\mu &=& \left( -gf^{abc}\omega^b_\mu\omega^c + 
(D_\mu\beta)^a \right){\delta\lambda}\nonumber\\
\delta\bar{\omega}^a_\mu &=& {1\over\eta}(\partial^\nu 
B_{\mu\nu}){\delta\lambda}\nonumber\\
\delta\beta^a &=& - 
gf^{abc}\beta^b\omega^c{\delta\lambda}\nonumber\\
\delta\bar{\beta}^a &=& - 
{1\over\zeta}\partial_\mu\bar{\omega}^{a\mu}{\delta\lambda}\nonumber\\
\delta\theta^a &=& \left(- gf^{abc}\theta^b\omega^c - 
\beta^a\right){\delta\lambda}\nonumber\\
\delta{\bar\theta}^a &=& 0.
\label{nabrst}
\end{eqnarray}
This set of transformations has the correct limits --- if $\omega^a$ is 
the only non-vanishing ghost, these would be the transformations 
corresponding to an $SU(N)$ symmetry, whereas if $f^{abc}$ and $C_\mu$ 
are set to zero, the abelian BRST transformations (\ref{abgbrs}) are 
recovered. It is straightforward to check that this set of 
transformations is 
nilpotent in a manner similar to the abelian case,
\begin{equation}
{\delta^2\over{\delta\lambda}^2}\left\{ A^a_\mu, B^a_{\mu\nu}, 
C^a_{\mu\nu}, 
\omega^a, \omega^a_\mu, \beta^a, \theta^a \right\} = 0.
\label{nabsq}
\end{equation}
It is also straightforward to show that the set of the BRST 
transformations as posited above leaves the sum of the gauge-fixing and 
ghost Lagrangians invariant,
\begin{equation}
{\delta\over{\delta\lambda}}({\cal L}_{gf} + {\cal L}_{FP}) = 0.
\end{equation}
The total BRST-invariant action can now be written as a 
sum of three terms, the gauge term, the gauge fixing term, and the 
ghost contribution,
\begin{equation}
S = \int d^4x ({\cal L}_0 + {\cal L}_{gf} + {\cal L}_{FP}),
\label{nabeff}
\end{equation}
with 
\begin{eqnarray}
{\cal L}_0 &=& - {1\over 4}F_{\mu\nu}^aF^{a\mu\nu} - {1\over 
12}\widetilde 
H_{\mu\nu\lambda}^a\widetilde H^{a\mu\nu\lambda} + 
{m\over
4}\epsilon^{\mu\nu\rho\lambda}B_{\mu\nu}^aF_{\lambda\rho}^a,\nonumber\\
{\cal L}_{gf} &=& - {1\over {2\xi}}(\partial_\mu A^{a\mu})^2 - 
{1\over{2\eta}}(\partial_\nu B^{a\mu\nu})^2 - 
{1\over\zeta}(\partial_\mu\bar{\omega}^{a\mu})(\partial_\nu\omega^{a\nu
}), 
\nonumber\\ 
{\cal L}_{FP} &=&\ \partial^\mu\bar{\omega}^a(D_\mu\omega)^a - 
gf^{abc}\partial^\nu\bar{\omega}^{a\mu}B^b_{\mu\nu}\omega^c + 
\partial^\nu\bar{\omega}^{a\mu}(D_{[\mu}\omega_{\nu]})^a \nonumber\\
&&\qquad\qquad\qquad\qquad
- gf^{abc}\partial_\nu\bar{\beta}^a\omega^{b\nu}\omega^c + 
\partial^\mu\bar{\beta}^a(D_\mu\beta)^a.
\label{nabrstac}
\end{eqnarray}
This action is fully gauge fixed with respect to the $SU(N)$ gauge 
transformations, as well as the vector gauge transformations 
(\ref{nabkr}), but it is invariant under the BRST transformations 
given in (\ref{nabrst}). This action also implies the nilpotence of 
the BRST transformation on $\bar{\omega},\bar{\omega}_\mu,\bar{\beta}$ 
and 
$\bar\theta$,
\begin{equation}
{\delta^2\over{\delta\lambda}^2}\left\{\bar{\omega}^a, 
\bar{\omega}^a_\mu, \bar{\beta}^a, 
\bar\theta^a \right\} = 0,
\label{nabarsq}
\end{equation}
taking into account their equations of motion. Off shell, their third 
variations vanish,
\begin{equation}
{\delta^3\over{\delta\lambda}^3}\left\{\bar{\omega}^a, 
\bar{\omega}^a_\mu, \bar{\beta}^a, 
\bar\theta^a \right\} = 0,
\label{nabcu}
\end{equation}
just as in the case of $\bar{\omega}^a$ in the case of pure Yang-Mills 
theory. It is now possible to construct the BRST invariant Noether 
current for this action in the same manner as in the abelian case. The 
variation of the action vanishes,
\begin{equation}
{\delta\over{\delta\lambda}}S = \int d^4x \partial_\mu Y^\mu = 0,
\end{equation}
with 
\begin{eqnarray}
Y^\mu = {m\over 2} \epsilon^{{\mu\nu\lambda}}\omega^a_\nu 
F^a_{\lambda\rho} - 
{1\over\xi}(\partial_\nu A^{a\nu}(D^\mu\omega)^a &-& 
{1\over\eta}(\partial^\lambda 
B^a_{\nu\lambda})\left(gf^{abc}B^{b\mu\nu}\omega^c 
- (D^{[\mu}\omega^{\nu]})^a + 
gf^{abc}F^{b\mu\nu}\theta^c\right)\nonumber \\
&-& {1\over\zeta}(\partial_\lambda\bar{\omega}^{a\lambda})( - 
gf^{abc}\omega^{b\mu}\omega^c + (D^\mu\beta)^a).
\end{eqnarray}
The Noether current is therefore
\begin{eqnarray}
j^\mu = &&\sum{{\delta {\cal 
L}}\over{\delta\partial_\mu\phi}}{{\delta\phi}
\over{{\delta\lambda}}} - Y^\mu \nonumber\\ 
= &&\left( - F^{a\mu\nu} + {m\over 2}\epsilon^{{\mu\nu\lambda}\rho} 
B^a_{\lambda\rho} - {1\over\xi}g^{\mu\nu}(\partial_\lambda 
A^{a\lambda}) - 
gf^{abc}C^b_\lambda H^{c{\mu\nu\lambda}}\right)(D_\nu\omega)^a 
\nonumber\\ && - 
{1\over 2}H^{a{\mu\nu\lambda}}\left( - gf^{abc}B^b_{\nu\lambda}\omega^c 
 + 
(D_\nu\omega_\lambda - D_\lambda\omega_\nu)^a - 
gf^{abc}F^b_{\nu\lambda}\theta^c\right) \nonumber\\ && + 
{1\over\eta}(\partial^\lambda B^a_{\nu\lambda})\left( - 
gf^{abc}B^{b\mu\nu}\omega^c + (D^\mu\omega^\nu - 
D^\nu\omega^\mu)^a\right) + 
{1\over{\zeta\eta}}(\partial_\lambda\omega^{a\lambda})(\partial^\sigma 
B^{a\mu\sigma}) \nonumber\\
&&- {1\over 2}gf^{abc}(\partial^\mu\bar{\omega}^a)\omega^b\omega^c - 
(\partial^\mu\bar{\omega}^{a\nu} - 
\partial^\nu\bar{\omega}^{a\mu})\left( - 
gf^{abc}\omega^b_\nu\omega^c + (D_\nu\beta)^a\right)\nonumber\\
&& + gf^{abc}\partial^\mu\bar{\beta}^a\beta^b\omega^c - 
{1\over\zeta}(\partial_\lambda\bar{\omega}^{a\lambda})\left( - 
gf^{abc}\omega^{b\mu}\omega^c + (D^\mu\beta)^a\right).
\label{nabcur}
\end{eqnarray}

Just as in the abelian case, the BRST charge constructed from this 
current,
\begin{equation}
Q_{BRST} = \int j^0\ d^3x 
\end{equation}
is nilpotent, $Q^2_{BRST} = 0$, and implements the BRST transformations 
on the fields, as can be explicitly checked by writing out the charge 
in terms of the canonically conjugate momenta to the fields and the 
ghosts.

\section{Discussion}

Let me first summarize what has been done so far. First I constructed a 
BRST-invariant gauge fixed action for the abelian mass generation 
mechanism. The transformations in the abelian case were then 
generalized to the non-abelian mechanism. The non-abelian BRST 
transformations reduce to those for the abelian case or for the free 
Yang-Mills case in the appropriate limits. The gauge fixed effective 
Lagrangian was constructed by including the appropriate ghost terms 
which leave the total action invariant under the BRST transformations. 
This invariance leads to a conserved BRST charge which is nilpotent on 
the Fock space. The cohomology of the BRST charge can be identified 
with the physical subspace of the Hilbert space, and the unitarity of 
the S-matrix is guaranteed on the physical states.

It is possible to compute the Slavnov-Taylor identities for the 
non-abelian theory starting from the BRST-invariant effective action of 
(\ref{nabeff}). It is outside the scope of this paper to do that, or to 
construct counterterms and prove perturbative renormalizability of the 
theory, which will be done elsewhere. It should be noted that no 
kinetic term (or any other quadratic term) for $C_\mu$ was required for 
the nilpotence of the BRST transformations, i.e., for the construction 
of a BRST-invariant quantum action for the theory. Thus $C_\mu$ remains 
a non-dynamical auxiliary field at tree level even after quantization.

Does anything change when fermions are coupled to the theory? If the 
fermions are minimally coupled {\em only} to the gauge field $A_\mu$, 
it is easy to see that the resulting theory can be made BRST-invariant 
in the same way as before after adding in the usual BRST 
transformations of fermions in gauge theories. In the abelian model, 
fermions cannot couple to the antisymmetric tensor because the minimal 
coupling breaks the vector gauge symmetry. In the non-abelian model, 
the vector gauge symmetry is enforced by the introduction of the 
auxiliary $C_\mu$. As a result it is possible to couple the non-abelian 
antisymmetric tensor to fermions, the general term for minimal coupling 
being $\bar\psi(a + b\gamma_5)\sigma^{\mu\nu}(B_{\mu\nu} - 
D_{[\mu}C_{\nu]}\psi$. This term is invariant under both the continuous 
symmetries, but breaks chiral symmetry. It is plausible that fermion 
mass is generated as a dynamical effect as a result of chiral symmetry 
breaking via this term.

\centerline{\bf Acknowledgements}

This work was supported by a grant from the Particle Physics and 
Astronomy Research Council at the University of Sussex at Brighton.





\begin{thebibliography}{99}
\baselineskip=23pt

\bibitem{opd}{V. I. Ogievetskii and I. V. Polubarinov, 
{\sl Sov. J. Nucl. Phys.}
({\sl Iad. Fiz.}) {\bf 4} (1967) 156; S. Deser, 
{\sl Phys. Rev.} {\bf 187} (1969) 1931.}
\bibitem{aurtak}{A. Aurilia and Y. Takahashi, {\it Prog. Theor. Phys.} 
{\bf 66} (1981) 693.}
\bibitem{trg}{T. R. Govindarajan, {\it J. Phys. G} {\bf 8} (1982) L17.}
\bibitem{abl}{T. J. Allen, M. J. Bowick and A. Lahiri, 
{\it Mod. Phys. Lett.} {\bf A6} (1991) 559.}
\bibitem{minwar}{J. A. Minahan and R. C. Warner, {\it St\" uckelberg 
Revisited}, Florida U. Preprint UFIFT-HEP-89-15.}






\bibitem{collins}{J. C. Collins, {\sl Renormalization}, Cambridge 
University Press, 1984.}
\bibitem{gvm} {A. Lahiri, {\it Generating vector boson masses},  
Los Alamos preprint LA-UR-92-3477,  hep-th/9301060.}
\bibitem{fretow}{D. Z. Freedman and P. K. Townsend, {\sl Nucl. Phys.} 
{\bf B177} (1981) 282.}
\bibitem{brs}{C. Becchi, A. Rouet and R. Stora, {\sl Phys. Lett.} {\bf 
B32} (1974) 344; {\sl Comm. Math. Phys.} {\bf 42} (1975) 127; {\sl Ann. 
Phys.} {\bf 98} (1976) 287.}
\bibitem{tyu} {I. V. Tyutin, Lebedev Institute preprint LEBEDEV-75-39, 
(1975), unpublished.}
\bibitem{bec}{C. Becchi, preprint GEF-TH-96-10, hep-th/9607181.} 
\bibitem{thibau} {J. Thierry-Mieg and L. Baulieu, {\sl Nucl. Phys.}
{\bf B228} (1983) 259.}
\bibitem{hentei} {M. Henneaux and C. Teitelboim, {\sl Quantization of 
Gauge Systems}, Princeton University Press, 1992.}
\bibitem{abcon} A. Lahiri, {\sl Mod. Phys. Lett.} {\bf A8} (1993) 2403.
\bibitem{nabcon} {T. J. Allen and A. Lahiri, {\it in preparation}.}
\bibitem{hwalee}{D. S. Hwang and C.-Y. Lee, {\it Nonabelian Topological 
Mass Generation in 4 Dimensions}, hep-th/9512216.}

\end{thebibliography}
\end{document}